\documentclass[fleqn,usenatbib]{mnras}

\usepackage{newtxtext,newtxmath}

\usepackage[T1]{fontenc}

\usepackage{graphicx}	
\usepackage{amsmath}	

\title[Correlated Current Variations]{Upper Limit on Correlated Current Variations in the Crab Pulsar}

\author[M. Vivekanand]{
M. Vivekanand\thanks{E-mail: viv.maddali@gmail.com}
\\
No. $24$, NTI Layout $1$\textsuperscript{st} Stage, $3$\textsuperscript{rd} Main,
$1$\textsuperscript{st} Cross, Nagasettyhalli, Bangalore $560094$, India.
}

\date{Accepted XXX. Received YYY; in original form ZZZ}

\pubyear{2023}

\begin{document}
\label{firstpage}
\pagerange{\pageref{firstpage}--\pageref{lastpage}}
\maketitle

\begin{abstract}
   The high energy emission of rotation powered pulsars is supposed to be produced in "gaps" 
   in the pulsar magnetosphere where charges are accelerated and currents are produced. The 
   rest of the magnetosphere is supposed to be mostly a "force-free" plasma without any 
   currents. Two important currents are the main current that flows away from the pulsar, 
   that produces the observed radiation, and the current that returns to the pulsar to 
   maintain charge neutrality. This work attempts to study the return current in the Crab 
   pulsar using the soft X-ray data from the {\it{NICER}} observatory. It is assumed that the 
   two currents vary as a function of time. This would modulate the electric fields in the 
   "gaps", which would affect the observed X-ray flux. These flux variations will show up 
   only in the on-pulse phases, while those caused by the Crab Nebula, instrumental effects, 
   etc. will be present in the off-pulse phases also. This work obtains the correlation 
   coefficient of the flux variations in the two peaks of the Crab pulsar, after removing the 
   off-pulse flux variations. No correlation was observed; its error of $0.000012$ sets an upper 
   limit of $0.036\%$ on the rms variation of correlated X-ray flux in the Crab pulsar. Reasons 
   exist for the return current variations to be correlated, while the main current variations 
   are probably uncorrelated. So the above number is considered an upper limit 
   on correlated return current variations, which may be an important constraint for pulsar 
   magnetospheric structure.
\end{abstract}

\begin{keywords} 
Stars: neutron -- Stars: pulsars: general -- Stars: pulsars: individual PSR J0534+2200 -- Stars: pulsars: individual PSR B0531+21 -- Stars: pulsars: individual PSR B0823+26 -- Stars: pulsars: individual PSR B0943+10 -- Stars: pulsars: individual PSR B1822-09 -- X-rays: general --
\end{keywords} 
%

\section{Introduction}

Rotation powered pulsars (RPPs) have intense magnetic fields whose rotation causes
intense electric fields outside the pulsar. These pull out electrons and ions from 
the surface of the pulsar which are then accelerated and produce electron-positron
pairs. These pairs are themselves accelerated and produce further pairs. Eventually 
this cascade leads to the outside of the pulsar being filled with a plasma of 
electrons, positrons and ions co-rotating with the pulsar -- this is known as the 
magnetosphere. For original contributions to this subject see \cite{Goldreich1969},
\cite{Ostriker1969}, \cite{Sturrock1971}, \cite{Ruderman1975}, \cite{Arons1979},
\cite{Arons1983} and \cite{Cheng1986}.

Setting aside pair production for a moment, consider what happens when a current of, 
say, electrons leaves the pulsar. If no other physics intervenes, then a positive
charge builds up on the pulsar as a function of time. This would reduce the 
accelerating electric field and eventually inhibit the outward electron current. The 
pulsar may end up as a charged and magnetized globe rotating at its period, but not
producing the high energy X-rays and $\gamma$-rays that are emitted by the Crab pulsar. 
This would be an inert electrosphere \citep{Arons2009, Spitkovsky2011}. Clearly pair 
production intervenes -- the accelerated charges emit curvature or synchrotron or 
inverse Compton photons of high energy, which form the observed radiation. These 
photons also produce electron-positron pairs in the strong magnetic field of the 
pulsar. The electrons of the pairs are accelerated in the same direction as the 
original current, i.e., away from the pulsar, while the positrons are accelerated in 
the reverse direction, and can reach the pulsar through a region of "force-free" 
plasma. But this can not be the return current -- it has the wrong sign of charge. 
Therefore a proper return current is a vital component of the pulsar high energy 
emission mechanism. See \cite{Contopoulos1999}, \cite{Cheng2011}, \cite{Hirotani2011}, 
\cite{Spitkovsky2011}, \cite{Arons2011}, \cite{Petri2011} and \cite{Contopoulos2019} 
for details of the pulsar magnetosphere and the two currents and their relation to 
the high energy emission mechanism. To the best of my knowledge the return current 
was first discussed seriously by \cite{Arons1979}. 

A brief summary of the essential features of the two currents is given below. It is 
an observer's perspective of the essential features of the currents in a RPP, Several 
theoretical details are unimportant for the current purpose and are therefore ignored 
(see \cite{Cerutti2016} and \cite{Philippov2020} for illustration).

\subsection{The main and return currents in a RPP}

\begin{figure}
\centering
\advance\leftskip-0.3cm
\includegraphics[keepaspectratio=true,scale=1.0,width=0.52\textwidth]{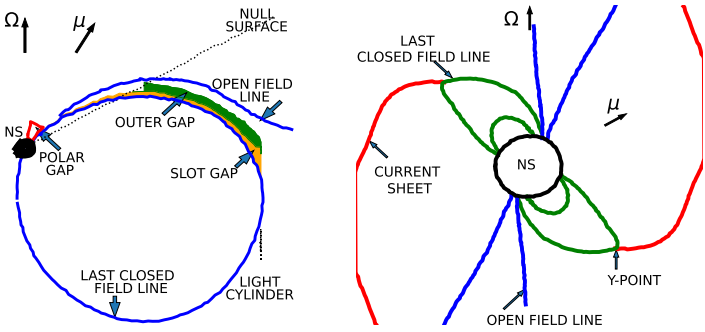}
\vskip-0.25cm
\caption{
	(a) Left panel: Cartoon of Fig. $1$ of \protect \cite{Hirotani2011}. (b) Right panel: Cartoon of 
	Fig. $2$ of \protect \cite{Spitkovsky2006}. These cartoons are intended only for convenience. For 
	any scientific details the original figures must be referred to. NS is the neutron star 
	(RPP); it is not to scale with respect to the radius of the light cylinder. $\Omega$ 
	and $\mu$ are its rotation and magnetic dipole axes.
	}
\label{fig1}
\end{figure}

Figure~\ref{fig1}~(a) shows a cartoon of the pulsar magnetosphere and its
various "gaps". It is a two dimensional projection of a three dimensional object.
The RPP is defined by its rotation axis $\Omega$ and magnetic dipole axis $\mu$. 
The magnetosphere is divided into open and closed magnetic field line regions; 
the field line labeled "last closed field line" defines the boundary between the 
two. A charge in the closed field line region can not leave the magnetosphere. 
The rest of the magnetosphere shown consists of open filed line region in which 
currents can leave and return to the pulsar.

In the open field line region there are three gaps. The polar cap gap lies just 
above the surface of the pulsar; it was the earliest gap proposed and is currently 
believed to be the source of the radio radiation of the RPP. It was found unsuitable 
for the high energy emission because of high magnetic opacity for photons near the 
surface of the pulsar. So the outer gap was proposed; it extends from a so called 
null surface at one end to the light cylinder at the other end, both shown as dotted 
lines in Fig.~\ref{fig1}~(a), and has a thickness much smaller than its length. Its 
lower boundary is quite close to the last closed field line; this is the green area 
in the figure.

Between the lower boundary of the outer gap and the last closed field line lies the 
slot gap, a relatively thin gap shown in orange color in the figure. It extends right 
from the surface of the RPP to the light cylinder.

Now, the main current from the pulsar originates in these gaps and flows away from 
the pulsar; the charge carried away by this current depends upon whether the angle
between $\Omega$ and $\mu$ is acute or obtuse. It is not clear whether all three
gaps can operate simultaneously in the same RPP (polar cap gap and outer gap are 
believed to be mutually exclusive), but at least one of them is active. See 
\cite{Harding2011}, \cite{Harding2022} and references therein for details about 
these gaps. 

The return current reaches the RPP through a very narrow bundle of magnetic field 
lines along the last closed field line; this would be below the slot gap and probably 
almost coincides with what is known as the separatrix layer, which is like a narrow 
slot gap that extends beyond the light cylinder. For details of the return current 
please refer \cite{Arons2011} and particularly its Fig.~$5$, and Fig~$15.2$ of 
\cite{Arons2009}, and figures $1$ and $2$ of \cite{Contopoulos2019}.

The structure of both currents is a strong function of the angle $\alpha$ between the
the rotation axis $\Omega$ and the magnetic axis $\mu$. 

\subsection{Some properties of the main current}

The properties of the main current obviously depend upon the gap involved. For the
polar cap gap early research postulated periodic build up and breakdown of the gap
on timescales of micro seconds \citep{Ruderman1975}, leading to a current that 
sparks on time scales of micro seconds. This was revised later to a 
space charge limited quasi steady state current due to the work function of iron 
ions on the surface of the neutron star. The main current in all gaps depends
upon the so called "favorably curved magnetic field lines" (see \cite{Cheng2011} 
and references therein). In all gaps pair production reduces the effective electric 
field due to screening by the pairs -- then one has a space charge limited flow. 
In the outer gap the current is quite low in the accelerating region but much 
larger in the screening region \citep{Cheng2011}. The magnitude of the main 
current depends upon the gap thickness. So high-altitude slot gaps can not produce 
sufficient high energy flux due to their being thin \citep{Hirotani2011}.

\subsection{The return current and the current sheet}

The last closed field line has been depicted as almost a circle in Fig.~\ref{fig1}~(a).
Modern particle-in-cell simulations show it to be actually that depicted in 
Fig.~\ref{fig1}~(b).  It develops a sharp projection at the point where it just touches 
the light cylinder; this is known as the "Y-point". Here it also connects with the so 
called "current sheet", shown in red in the figure (strictly the Y-point does not 
necessarily have to touch the light cylinder \citep{Spitkovsky2011}). Most of the return 
current originates far away from the RPP and flows towards it in the current sheet. At 
the Y-point the return current splits into two streams along the last closed field line, 
each proceeding to one of the two magnetic poles of the RPP. Although the current sheet 
has been represented by a red curve of uniform width in Fig.~\ref{fig1}~(b), it is 
thickest at the Y-point and reduces in thickness further away from the RPP.

\subsection{Some properties of the return current}

The above is a two dimensional projection of a three dimensional object, the projection
being in the plane containing the vectors $\Omega$ and $\mu$. So the current 
sheet is actually a current ring or a current torus, and the Y-point is actually a 
Y-volume since it has finite extension in all three dimensions, and the return current 
is flowing down the edge of a three dimensional magnetic funnel at the polar caps of the 
RPP. The nature of the current sheet and the return current it provides depend upon the 
angle $\alpha$. The magnitude of the return current decreases, and its magnetospheric 
distribution changes dramatically, with $\alpha$ \citep{Spitkovsky2011}. The return 
current path is also expected to carry a counter-streaming current \citep{Arons2011, 
Contopoulos2019}. In this work we are concerned with the net return current which is 
the sum of all counter-streaming currents, if at all they exist.

It is now suspected that the high energy emission from RPPs originates due to reconnection 
in the current sheet, presumably at the Y-point. See \cite{Spitkovsky2011}, \cite{Arons2011} 
and \cite{Petri2011} for details.

\subsection{The philosophy of this analysis}

To begin with, it is assumed that both the main and return currents together determine the 
maximum available electric potential in the gaps, which is related to the X-ray flux emitted
by the RPP. The gaps can be thought of as electrical batteries that are charged and 
discharged by the two currents in a collective manner, implying higher and lower electric 
potential available for particle acceleration, respectively. However the dynamics of the two 
currents 
are likely to be vastly different -- the main current is powered by the immense rotational 
inertia of the pulsar and its intense magnetic field, and depends upon the details of the 
gaps, the pair production process, and the high energy emission mechanism. On the other hand, 
the return current begins its journey somewhere far away from the RPP and depends critically 
upon the details of the current sheet and the pulsar magnetosphere.

Next it is assumed that both currents vary as a function of time. Clearly a steady, 
unchanging current is not logical, given the highly energetic and almost explosive 
environment in the pulsar magnetosphere. What is not known is the time scale 
of such variations. It is assumed here that all possible time scales of variation exist 
in the currents.

Next, one notes regarding the main current that ($1$) its variations are unlikely to be 
correlated at the two poles, since the emission at each pole is expected to be 
independent, and ($2$) the emission at each pole may or may not be correlated across
different phases in the folded light curve (FLC) of the RPP, which are equivalent to
different regions of emissions in the gaps. On the other hand, variations imposed upon
the return current beyond the Y-point are certainly correlated not only at the 
two poles but also at all phases in the FLC, although one can not rule out 
de-correlating variations being imposed on the return current between the Y-point 
and the surface of the RPP. 

The purpose of this work is to study correlated flux variations if any in the Crab 
pulsar. The basic premise of this work is that the return current variations (beyond 
the Y-point) should be imprinted on the X-ray flux of the RPP, while the main current 
variations may or may not be imprinted. Further, these variations will exist in the 
pulsar flux, and not in that from the nebula. So the technique used here is to
divide the FLC of the Crab pulsar into three regions of phase -- the main peak and 
the second peak (forming the so called on-pulse region), and the off-pulse region. 
The nebular flux and instrumental effects will exist at all phases 
while the current variations are expected only in the on-pulse. The idea is 
to correlate the X-ray flux in the two peaks of the Crab pulsar, after estimating 
the nebular flux variations and/or instrumental effects from the off-pulse phases and
removing them from the on-pulse phases. At best this number will be determined only 
by the return current variations; however one can not rule out correlated main 
current variations altogether.

Incidentally, variations of currents in a RPP can also manifest as variations of
electromagnetic torque on the RPP, leading to fluctuations in its rotation period.
This is commonly known as timing noise. This work
addresses a particular component of it, that which is correlated across the FLC of 
the RPP. As mentioned earlier in this section, only the return current may display
such a correlation. Given the possibility of de-correlating effects before the Y-point,
it is possible that the correlated return current variations may be a very small 
fraction of the total current variations in the RPP. In such a situation the correlated
return current variations may not leave a measurable imprint on the timing noise of
a RPP.

In footnote $6$ on page $387$ \cite{Arons2009} states that observational study of pulsar
currents has been an untouched subject. This work attempts to redress this issue.

\section{Observations and analysis}

\begin{figure}
\begin{center}
\advance\leftskip-0.3cm
\includegraphics[keepaspectratio=true,scale=1.0,width=0.52\textwidth]{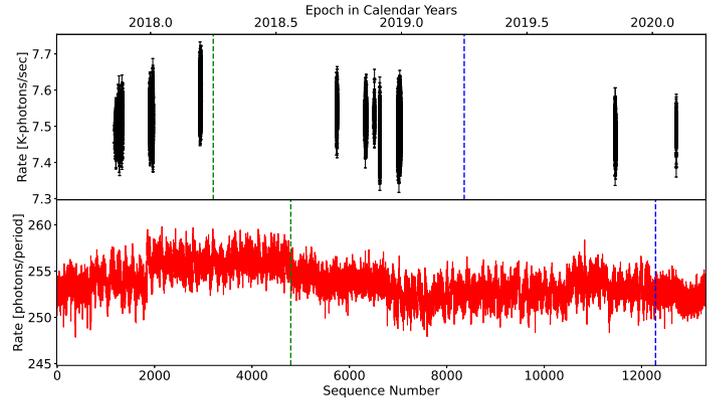}
\end{center}
\vskip-0.5cm
\caption{
	(a) Top panel: Long time light curve (LTLC) of the Crab pulsar from {\it{NICER}} 
	data in the energy range $1 - 10$ keV with a bin size of $10$ s, using the 
	$27$ observations specified in \protect \cite{Vivekanand2021}; the abscissa is exactly 
	the same as in its Fig.~$1$; the ordinate is in units of thousand photons per 
	second. (b) Bottom panel: LTLC using the same data but with a bin size of $300$ 
	periods, which is $\approx 10.125$ s; the abscissa is sequence number of the bins
	while the ordinate is in photons per period. The green and blue dashed lines mark 
	the same epochs respectively in both panels.
	}
\label{fig2}
\end{figure}

The details of the {\it{NICER}} observations used here and their preliminary analysis 
are given in \cite{Vivekanand2020, Vivekanand2021}.

The top panel of Fig.~\ref{fig2} is the long time light curve (LTLC) of the Crab pulsar
with a bin size of $10$ s in the energy range $1$ to $10$ keV. This consists of $137161$ 
s of data equivalent to $\approx 4.06$ million periods. It is similar to Fig.~$1$ of 
\cite{Vivekanand2021} except that a small amount of initial data is not included for
reasons stated there, and the energy range is different. The bottom panel of Fig.~\ref{fig2}
is the same data with a bin size of $300$ periods which is approximately $10.125$ s. The
binning is in number of periods because this analysis depends upon the phase in the FLC, 
so the period is the natural unit for binning. It also eliminates the error introduced 
by fractional periods of data at the beginning and the end of a time bin. By plotting 
the data in terms of the sequence number of the bins one eliminates the large gaps in the 
epoch in the top panel, leading to a better appreciation of the variability of the X-ray 
flux of the Crab pulsar. The green and blue dashed lines in the top panel represent the 
epochs $2018.25$ and $2019.25$ respectively. The corresponding lines in the bottom panel
mark the partitioning of the data in terms of these epochs. Thus the sequence numbers in
the bottom panel just before and just after the green dashed line correspond to the time 
bins just before and just after the green dashed line in the top panel. The same is
true for the blue dashed lined in both panels. However the correspondence is approximate
since the bins in both panels are slightly different in terms of time. Further, the
period bins in the bottom panel of Fig.~\ref{fig2} are slightly larger in terms
of time towards later epochs since the period of the Crab pulsar increases with epoch. 

The bottom panel of Fig.~\ref{fig2} was obtained by first searching for a new period at 
the start of each good time interval (GTI), then accumulating photon counts over $300$ 
contiguous periods for each bin. The incomplete bin at the end of the GTI is discarded.
The mean and rms of the counts are $253.79$ and $1.79$ respectively; the rms is a fraction
$1.79 / 253.79 = 0.00705$ of the mean or $\approx 0.71\%$. As shown later on this 
variation of X-ray flux is present at all phases in the FLC. Therefore this must be 
either due to the Crab nebula or due to instrumental effects such as pointing variations
of {\it{NICER}}. Now, this number is consistent with the $1^\prime$ pointing errors of 
{\it{NICER}} producing $2\% - 4\%$ flux variations.\footnote{https://heasarc.gsfc.nasa.gov/docs/nicer/data\_analysis/nicer\_analysis\_tips.html} These variations have to be removed from the data before 
estimating those that exist only in the on-pulse phases.

The analysis begins by defining three phase ranges in the period of the Crab pulsar --
first peak, second peak and off-pulse, labeled P1, P2 and P3 respectively, with respect 
to Fig.~$1$ of \cite{Vivekanand2022}. The three phase ranges were finalized after some
experimentation as $0.1015625$ to $0.40625$, $0.40625$ to $0.796875$, and the rest of 
the phase range, respectively, the specific numbers being chosen so as to correspond 
to integer multiples of $1/128$ phase, and also to maximize the off-pulse phase range 
without compromising on the on-pulse phase range. The final results of this work are
not sensitive to the exact choice of the above phase range limiters. P1 and P2 together 
form the on-pulse phase range while P3 is the off-pulse.

Throughout this work it will be assumed that component P3 contains no pulsar flux,
in spite of \cite{Tennant2001} discovering off-pulse non-thermal X-ray emission from 
the Crab pulsar at much softer X-ray energies, because their Fig.~$1$ shows that this
flux is about two orders of magnitude smaller than the flux at the first peak of
the Crab pulsar.

The next section implements the cross-correlation analysis upon the data of P1, P2 
and P3 to reproduce the $\approx 0.71\%$ flux variation in all three components. 
This is to validate our analysis method. In the following section these variations 
are estimated in the P3 component and removed from all three components. 
Cross-correlation of the modified P1 with the modified P2 should give us the 
correlated variations that exist only in the on-pulse, which will be attributed to 
current variations in the Crab pulsar. Correspondingly cross-correlation of the 
modified P1 and P2 with the modified P3 should result in almost zero correlation.

\section{Analysis of raw data}

\begin{figure}
\begin{center}
\advance\leftskip-0.3cm
\includegraphics[keepaspectratio=true,scale=1.0,width=0.52\textwidth]{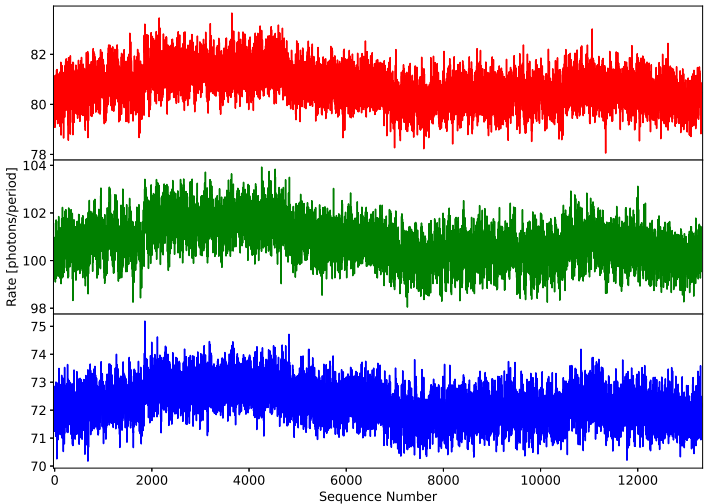}
\end{center}
\vskip-0.5cm
\caption{
	Same as the bottom panel of Fig.~\ref{fig2} but for each of the components
	P1, P2 and P3 of the FLC (top, middle and bottom panels respectively).
        }
\label{fig3}
\end{figure}

Figure~\ref{fig3} is the same as the bottom panel of Fig.~\ref{fig2} but for each of the 
components P1, P2 and P3 of the FLC; the sum of these three rates is exactly equal to 
the corresponding rate in the bottom panel of Fig.~\ref{fig2}. The flux variation of 
Fig.~\ref{fig2} exists in all three components of the FLC. It is know from Fig.~\ref{fig2} 
that the fractional flux variation is $\approx 0.71\%$ of the mean flux. This will now be 
derived by cross correlating the fluxes in the three components, so as to validate the 
procedure for the next section.

In Fig.~\ref{fig3} the X-ray flux data has been averaged over $300$ periods for convenience
of visualization and plotting. However the cross correlation of the data must be estimated
using the raw (un-averaged) data, since averaging the data before cross correlating it
biases it to larger values. This is because averaging the data decreases the denominator
while maintaining approximately the same numerator in the formula for cross correlation 
(Appendix A). 
In a larger context this problem is related to what is known as ``ecological fallacy"\footnote{https://en.wikipedia.org/wiki/Ecological\_fallacy} in which correlation of aggregate data is used to interpret 
possible correlation among individual data, which can be misleading. In our case 
therefore the cross correlation must be done with the data of count rate per a single 
period.

Table~\ref{tbl1} lists the estimated cross correlation $R$ between pairs of components in 
the FLC of the Crab pulsar, along with the derived fractional rms variation $\sigma_c$ 
of the correlated X-ray flux (see Appendix A, which also explains the error on the cross correlation).
The $\sigma_c$ values in Table~\ref{tbl1} and the upper and lower limits of errors 
have not been rounded off to bring out the fact that a uniform error of $0.0004$ or 
$0.0005$ is a good enough approximation instead of the upper and lower limits. 
Results for the three $\sigma_c$ values are $0.70(4)\%$, $0.66(5)\%$ and $0.69(4)\%$,
which are consistent with each other. Their weighted mean is $0.69(2)\%$ which is 
consistent with the value of $\approx 0.71\%$ estimated in the previous section.
Our analysis is therefore validated for use in the following section.

Appendix B discusses the so called ``ecological fallacy" and the variation of the 
cross correlation $R$ for data averaged over $M$ samples.

\begin{table}
\begin{center}
\caption{
	Cross correlation $R$ of X-ray flux data binned at $1$ (one) period between 
	pairs of phase components P1, P2 and P3. $\sigma_c$ is the resulting rms flux 
	variation of the correlated component as a fraction of the mean counts. The 
	error in the last digit in $R$ is given in brackets. See Appendix A for 
	computational details.
        } \label{tbl1}
\begin{tabular}{|c|c|c|}
\hline
\hline
Component pairs & $R$ (Correlation) & $\sigma_c$ (Fractional rms) \\
\hline
\hline
(P1, P2) & $R_{12} = 0.0044(5)$ & \Large $0.00701_{-0.00041}^{+0.00039}$ \\
\hline
(P1, P3) & $R_{13} = 0.0033(5)$ & \Large $0.00656_{-0.00052}^{+0.00048}$ \\
\hline
(P2, P3) & $R_{23} = 0.0041(5)$ & \Large $0.00691_{-0.00044}^{+0.00041}$ \\
\hline
\hline
\end{tabular}
\end{center}
\end{table}

\section{Analysis of filtered data}

In this section one will first estimate the smooth version of the flux in component
P3 (bottom panel of Fig.~\ref{fig3}), and use that to remove the corresponding flux
variation in components P1 and P2 before cross correlation. There are several 
algorithms for smoothing data; see any graduate school level text book on filtering 
data in the subject of digital signal processing. To ensure that the results of this 
section are independent of the smoothing algorithm, three filters were used: the 
Savitzky-Golay\footnote{https://en.wikipedia.org/wiki/Savitzky\%E2\%80\%93Golay\_filter} filter, 
the moving average filter\footnote{https://en.wikipedia.org/wiki/Moving\_average}, and the Wiener 
filter\footnote{https://en.wikipedia.org/wiki/Wiener\_filter\#:~:text=In\%20signal\%20processing\%2C\%20the\%20Wiener,noise\%20spectra\%2C\%20and\%20additive\%20noise.}.

\begin{figure}
\begin{center}
\advance\leftskip-0.3cm
\includegraphics[keepaspectratio=true,scale=1.0,width=0.52\textwidth]{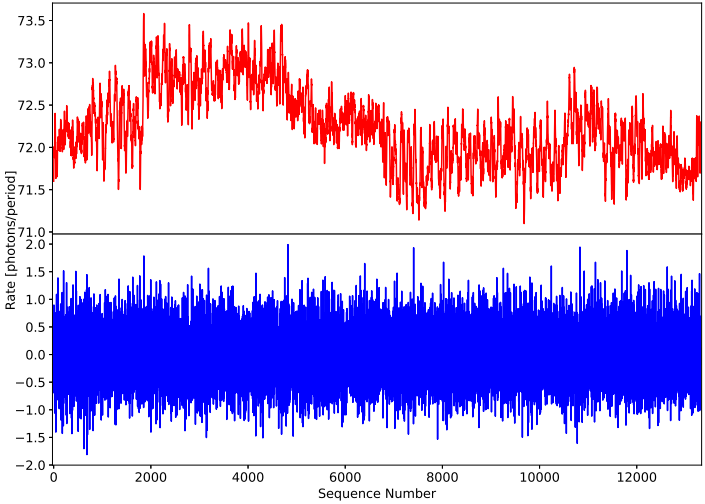}
\end{center}
\vskip-0.5cm
\caption{
	(a) Top panel: Data of phase component P3 (bottom panel of Fig.~\ref{fig3})
	filtered by a Savitzky-Golay algorithm using a window width of $30$ samples 
	and a polynomial of degree $2$. (b) Bottom panel: Difference between the 
	original and filtered data.
        }
\label{fig4}
\end{figure}

All three filters have one common parameter known as the the smoothing window $N_{sw}$ 
in units of data samples. In addition, the Savitzky-Golay filter uses a polynomial of
degree L, which is usually low ($\approx 2 - 3$). It smooths the data by representing 
it in each segment of $N_{sw}$ samples by a polynomial with L different coefficients. 
The moving average filter smooths the data by representing each data point by the 
average of $N_{sw}$ adjacent samples. The Wiener filter assumes that it knows the 
signal and noise power spectra, and applies low weightage to data segments (of width 
$N_{sw}$ samples) where the signal to noise ratio is low.

Fig.~\ref{fig4} shows an example of smoothing the data using the Savitzky-Golay filter
with $N_{sw} = 30$ and $L = 2$. The top panel displays the filtered data while the bottom 
panel displays the difference between the original and filtered data, which looks like 
white noise, as it should. Once again it is mentioned that although Fig.~\ref{fig4} 
shows X-ray flux data averaged over $300$ periods, the cross correlation is done using 
un-averaged data. Moreover, the cross correlation coefficient $R$ now depends upon the 
filter window width $N_{sw}$. The detail of this analysis are given in Appendix C and
the results are given in Table~\ref{tbl2}.

\begin{table}
\begin{center}
\caption{
	Cross correlation $R$ (in units of $10^{-5}$) of flux binned at $1$ (one) period 
	between pairs of phase components P1, P2 and P3, after estimating the flux 
	trend in P3 and removing it from all three components. Three filters were used 
	for estimating the trend in P3 -- Savitsky-Golay (SG), moving average (MA) and 
	Wiener (WI). See Appendix C for computational details.
        } \label{tbl2}
\begin{tabular}{|c|c|c|c|}
\hline
\hline
 & \multicolumn{3}{c}{Filter} \\
\hline
$R (\times 10^5) $ & SG & MA & WI \\
\hline
$R_{12}$ & $1.2 \pm 1.2$ & $0.40 \pm 0.59$ & $0.40 \pm 0.59$ \\
\hline
$R_{13}$ & $0.10 \pm 0.12$ & $-0.30 \pm 0.19$ & $-0.30 \pm 0.24$ \\
\hline
$R_{23}$ & $-0.2 \pm 0.19$ & $0.20 \pm 0.12$ & $0.20 \pm 0.12$ \\
\hline
\hline
\end{tabular}
\end{center}
\end{table}

Table~\ref{tbl2} gives the cross correlation coefficient $R$ for all three pairs of phase 
components for all three filters. Each $R$ in Table~\ref{tbl2} has been estimated with 
reference to a ``calibration'' $R$ that should ideally be zero; this is account for any
uncertainty in removing the smooth version of the flux of component P3 (see Appendix C 
for details of the analysis). All $R$ are either smaller than or comparable to their 
errors, so they are essentially zero. This is because they are the difference between
two numbers that have very similar values (Appendix C). Since no significant cross 
correlation has been estimated, the largest error on the cross correlations in 
Table~\ref{tbl2} ($1.2 \times 10^{-5}$) sets the upper limit for $\sigma_c$, which turn 
out to be $0.036\%$. Thus the correlated X-ray flux in the phase components P1 and P2 of 
the Crab pulsar is less than this value. In principle this includes variations from both 
the main and the return currents. Therefore this number is the upper limit on the 
correlated variation of the return current in the Crab pulsar.

\section{Correlation of X-ray flux from the two poles}

In section $1.5$ it was argued that the return current variations beyond the Y-point are 
correlated at the two poles. In this section attempt will be made to test this hypothesis
although it was acknowledged in section $1.5$ that variations imposed upon the return
current before the Y-point may lead to de-correlation.

The observed X-ray radiation from a RPP is emitted in the gaps in the pulsar 
magnetosphere. Within the gaps the radiation is emitted with its momentum vector tangent 
to the local magnetic field line; this radiation is observed whenever the line of sight to 
the RPP is parallel to this tangent. Now in Fig.~\ref{fig1}~(a) only one half of the
magnetosphere is drawn -- a symmetrical gap system exists at the other magnetic pole.
Therefore it is possible that one may observe radiation from both poles depending upon
the angle $\alpha$ between the rotation and magnetic axis, and the angle between the
rotation axis and the line of sight. It is easier to visualize this in Fig.~\ref{fig1}~(b)
where magnetic field lines are drawn at both poles. Now, there will be a delay in the 
arrival time of radiation from the pole farther away from the observer, which would be 
typically $\approx$ a light cylinder distance farther away from the closer pole. This 
distance would take the radiation about $P/(2\pi)$ s of travel time where $P$ is the 
rotation period of the RPP; this number may be a bit larger depending upon the above
mentioned two angles. So the Crab pulsar's X-ray flux may have correlated flux 
variations with a delay of a fraction of the period, which would translate to an 
equivalent phase delay in the FLC.

This effect should be studied by using phase bins of much higher resolution than those
of the previous section. This would naturally decrease the number of photons in these 
bins making the estimation of cross correlation more difficult. In this section the 
FLC was divided into $128$ phase bins; the number of photons per phase bin (per period) 
reduced to typically $1 - 2$ photons, which is very small. The attempt was to cross 
correlate the data of 
the bin at the first peak of the FLC of the Crab pulsar, with all other bins except 
those in the off-pulse. Any correlated variations from the two poles should show up
as a secondary peak of cross correlation about $1/(2\pi)$ phase away from the first 
peak. As in the previous section, the variations in the off-pulse have to be removed
before correlating.

\begin{table}
\begin{center}
\caption{
	Cross correlation $R$ (in units of $10^{-5}$) of flux binned at $1$ (one) period 
	at phase bins $17$ and $69$ in the FLC with a resolution of $128$ bins; off-pulse
	(same P3 of the previous section) flux trend was removed from all three components.
	The Savitsky-Golay (SG) filter was used for trend estimation.
        } \label{tbl3}
\begin{tabular}{|c|c|}
\hline
\hline
 & Filter \\
\hline
$R (\times 10^5) $ & SG \\
\hline
\Large $R_{17, 69}$ & $0.4 \pm 0.1$ \\
\hline
\Large $R_{17, P3}$ & $-5.4 \pm 0.2$ \\
\hline
\Large $R_{69, P3}$ & $1.6 \pm 0.3$ \\
\hline
\hline
\end{tabular}
\end{center}
\end{table}

The analysis of the previous section was repeated using the Savitzky-Golay filter,
and the results are presented in Table~\ref{tbl3} which shows, for the purpose of
illustration, the result of correlating the data at phase bin number $17$ (at the 
first peak) and at bin number $69$ (at the second peak), using the same off-pulse 
phase range as earlier (P3). Although the correlations in the three rows are larger
than their formal errors, the correlations of data at bin numbers $17$ and $69$ 
with the data of P3 (bottom two rows of Table~\ref{tbl3}) are much larger than
the correlation of data of the two bins (top row of Table~\ref{tbl3}). This implies 
that the data trend of P3 has not been removed completely from the data of bins
$17$ and $69$. Clearly the correlations are consistent with no correlation, probably
because the number of photons per phase bin are too small for this exercise. 

\section{Discussion}

The summary of this work is that it attempts to estimate correlated soft X-ray flux 
variations in the on-pulse phases of the Crab pulsar, after removing flux variations 
due to the Crab nebula and/or instrumental effects. No correlation was detected; the 
measurement error on the correlation sets an upper limit of $0.036\%$ on the rms 
variation of correlated flux. In principle this could be due to both the main as 
well as the return currents; therefore the above number is an upper limit on the 
correlated variations of the return current in the Crab pulsar.

A few caveats will be stated before proceeding further. In this work it has been 
assumed that variations of the return current beyond the Y-point are carried back
to the star. However \cite{Arons2009} cautions that it is yet to be demonstrated
whether this is possible. Next, the return current has details that have been 
ignored in this work; for example $20\%$ of it flows to the pulsar along a 
different but close path; see \cite{Arons2009} for details.

In order to study the return current variations one should be able to decouple them
from the main current variations. This depends upon the time scales on which the
two are coupled, but currently this is not known. For illustration consider an
extreme example -- let the source of the return current be at half distance to the 
Crab nebula, which has a size of radius $\approx 2^\prime$ in the sky. Assuming that 
the Crab pulsar is at a distance of $\approx 2000$ parsec, the source of the return 
current is $\approx 1 / 60 \times \pi / 180 \times 2000 \approx 0.58$ parsec from 
the pulsar. So any change at the source of the return current will take at least 
$\approx 1.9$ years to be felt at the pulsar, and a consequent change in the 
main current will take another $\approx 1.9$ years to be registered back at the source 
of the return current. Under these circumstances variations in the two currents can 
be considered to be  decoupled and can be discussed separately, even though one 
does not know their relative strengths. On the other hand, if the above time 
scale is much shorter then variations in the two currents are closely coupled, and 
can not be discussed separately; they might form what is known in electronics as a 
feed back system.

An appropriate example here is a radio receiver with feed back -- a strong negative 
feed back will kill the output, a mild negative feed back will reduce the receiver 
noise power, a mild positive feed back will convert the receiver into an oscillator, 
and a strong positive feed back will push the receiver into saturation. The above 
should be kept in mind while discussing the return current.

One of the direct consequences of variations of the currents is the simultaneous
variation in observed emission at all energies. This would require a campaign of
multi-wavelength observations with high time resolution and high sensitivity, which 
are not available. However simultaneous observations at the radio and X-rays are
available on three pulsars --- PSR B0823+26 \citep{Sobey2015, Hermsen2018}, PSR 
B0943+10 \citep{Hermsen2013, Mereghetti2016} and PSR B1822-09 \citep{Hermsen2017}.
All three pulsars exhibit two main modes of emission -- a radio B (bright) mode 
in which the pulsars are bright in the radio, and a radio Q (quiescent) mode in 
which they are much weaker. Together these three RPPs make an interesting 
contradiction which the return current may or may not be able to resolve.

In PSR B0823+26 the X-ray flux in the B mode has a $\approx 20\%$ flux variation 
which the authors call a new kind of behavior \citep{Hermsen2018}. They speculate
that the Q mode is not a true "null" mode -- they believe some residual emission 
exists but being weak it is not observed at earth. They believe the mode changes 
in this pulsar are entirely different from those in the other two pulsars. They
believe that B0823+26 is accreting material from its external environment, either
from the interstellar medium, or from an accretion disk.

Now, any accreted material would be eventually ionized and must be funneled 
to the RPP along the same path as taken by the return current -- along other 
paths the currents are only allowed to leave the pulsar. In such a scenario, 
the return current can serve the same purpose as the accreted material, which 
will no longer be needed for the above model. Clearly several details have to 
be worked out in this scenario -- how much extra return current is required, 
can it be supplied by the current sheet, can the additional return current 
explain the observed spectral and temporal features as well as the accreted 
material can, and so on; but this possibility is worth exploring. 
\cite{Arons2009} mentions that fluctuations of the currents at the light 
cylinder can probably explain the ﬂickering nature of pulsars, although his
context may have been different.

If the return current is varying in PSR B0823+26 as speculated above, then 
one should also speculate if it is operating similarly in the other two pulsars 
also; here one encounters some difficulties.

The soft X-ray emission is correlated with the radio emission in B0823+26 -- 
in the B mode both radio and X-rays are observed, while both are not observed
in the Q mode. The return current can offer a simple explanation for this --
it is larger in the Q mode than in the B mode, so the maximum electric
potential available for particle acceleration in the gaps is smaller in the
Q mode. Since this affects the basic emission mechanism of the RPP, radiation
at all wavelengths can be expected to decrease.

However, the soft X-ray emission is anti-correlated with the radio emission in 
PSR B0943+10 -- the radio emission is weaker in the Q mode while the X-ray
emission is weaker in the B mode  by a factor of $\approx 2$. Clearly the same 
return current that can
explain the broad band behavior of B0823+26 will fail in the case of B0943+10.
One can attempt to salvage the situation by noting that the return current
behavior can be different in the two pulsars because B0943+10 is a nearly 
aligned rotator ($\alpha \approx 0^\circ$) while B0823+26 is a nearly 
orthogonal rotator ($\alpha \approx 90^\circ$), and the return current is a 
strong function of the angle $\alpha$ between the rotation and the magnetic 
axis. Even then to explain the anti-correlation a new element has to brought 
in -- formation of coherent bunches of particles.

Suppose that in PSR B0943+10 a high return current causes a higher number of
coherent bunches of particles; this would imply a higher level of radio 
emission even though the maximum available electric potential has decreased.
But this is not sufficient -- one has to come up with a reason why this
effect does not operate in PSR B0823+26. Clearly one is reaching the limits 
of reasonable speculation. This point is further emphasized by the fact that 
in PSR B1822-09, which is an orthogonal rotator, the radio and X-ray emission
are uncorrelated.

To summarize the discussion so far, the return current can explain (in principle)
the behavior of PSR B0823+26, but fails to explain the behavior of PSR B0943+10
and PSR B1822-09. See the references above for much greater details in the 
behavior of these RPPs. It is not clear what causes these behavior, but the 
return current may be a partial explanation.

So far one focused on what may be called the traditional physics of RPPs, in which
the gaps are the sites of pair production as well as emission of observed radiation.
Recent research favors the observed radiation to be produced outside the light 
cylinder. It involves concepts such as the striped wind in which the magnetic 
polarity is continuously reversed, relativistic beaming effects and so on; it 
depends upon the inverse Compton scattering process. This model is apparently 
successful in fitting the observed $\gamma$-ray FLCs of the Geminga and Vela RPPs, 
as well as the optical polarization data of the Crab pulsar (see \cite{Petri2011} 
and references therein).

It is possible that in such models the role of the return current is merely to
maintain sufficient electrical potential in the gaps for pair production.
However in this new scenario particle acceleration can occur in the current
sheet near the light cylinder by the method of magnetic reconnection 
\citep{Kirk2009, Arons2011}. In these models the gaps are no longer the only
sites of particle acceleration. So in these models maybe the return current 
has the simplest role -- that of ensuring that the pulsar does not end up as 
an inert electrosphere.

\section{Data availability}

The data underlying this article are available at the
NICER\footnote{https://heasarc.gsfc.nasa.gov/docs/nicer/} observatory site at HEASARC
(NASA).

\section{Acknowledgments}
I thank Craig Markwardt for discussion on some technical issues such as pointing errors. 
I thank I. Contopoulos for discussion regarding the return current.

\appendix

\section{Formula for $R$ and $\sigma_c$}

Let $X_i, Y_i$ ($i = 1, ..., N$) be N samples of data from two different statistical 
distributions that have some correlation. Let their mean values be $<X> = \left ( 
\sum_i^N X_i \right ) / N$ and $<Y> = \left ( \sum_i^N Y_i \right ) / N$ and their 
standard deviations be $\sigma_X = \left < \left ( X - <X> \right )^2 \right >$ and 
$\sigma_Y = \left < \left ( Y - <Y> \right )^2 \right >$ respectively.  Their cross 
correlation is defined as
\begin{equation}
R_{XY}  = \left < \left ( X - <X> \right ) \left ( Y - <Y> \right ) \right >
                  / (\sigma_X \sigma_Y).
\end{equation}
The standard deviation of the correlation coefficient is $(1 - R_{XY}^2) / \sqrt{N}$
\citep{Bowley1928}\footnote{https://www.jstor.org/stable/2277400?seq=1\#page\_scan\_tab\_contents}.
Since the correlation coefficient here is very small compared to $1$, 
the error on $R_{XY}$ has been estimated as $1 / \sqrt{N}$ in column $2$ of Table~\ref{tbl1},
with $N \approx 4.06$ million.

Now let the two data consist of a correlated signal $c_i$ and uncorrelated noise $s_i$, 
both defined as fractions of the mean value, so that
\begin{equation}
\begin{array}{ll}
X_i  & = <X> \left ( 1 + c_i + s_{Xi} \right ) \\
Y_i  & = <Y> \left ( 1 + c_i + s_{Yi} \right ). \\
\end{array}
\end{equation}
The correlated signal $c_i$ is common to both data while the noise in the two data streams 
are different and therefore uncorrelated, not only among themselves but also with the 
correlated signal. The statistical properties of the correlated signal are:
\begin{equation}
\begin{array}{ll}
<c> = 0; \sigma_c^2 = \left < \left ( c - <c> \right )^2 \right >  = \left < c^2 \right > \\
<c s_X> = <c s_Y> = 0. \\
\end{array}
\end{equation}
The statistical properties of the noise signals are:
\begin{equation}
\begin{array}{ll}
<s_X> = 0; \sigma_{sX} = \left < \left ( s_X - <s_X> \right )^2 \right >  = \left < s_X^2 \right > \\
<s_Y> = 0; \sigma_{sY} = \left < \left ( s_Y - <s_Y> \right )^2 \right > = \left < s_Y^2 \right > \\
<s_X s_Y> = 0. \\
\end{array}
\end{equation}
Now the noise signals obey Poisson statistics so their variance is equal to the mean counts. So
the variance of the product $(<X> s_X)$ is $<X>$, so that
\begin{equation}
\begin{array}{ll}
<X>^2 \sigma_{sX}^2 = <X> \Rightarrow \sigma_{sX} = 1/\sqrt{<X>} \\
<Y>^2 \sigma_{sY}^2 = <Y> \Rightarrow \sigma_{sY} = 1/\sqrt{<Y>} \\
\end{array}
\end{equation}
Therefore
\begin{equation}
\begin{array}{ll}
R_{XY}  &= \left < \left ( c + s_X \right ) \left ( c + s_Y \right ) \right >
		  / \left ( \sqrt{\sigma_c^2 + \sigma_{sX}^2} \sqrt{\sigma_c^2 + \sigma_{sY}^2} \right ) \\
        &= \sigma_c^2 / \left ( \sqrt{\sigma_c^2 + \sigma_{sX}^2} \sqrt{\sigma_c^2 + \sigma_{sY}^2} \right ) \\
	&= 1 / \left ( \sqrt{1 + (\sigma_{sX}/\sigma_c)^2} \sqrt{1 + (\sigma_{sY}/\sigma_c)^2} \right ) \\
\end{array}
\end{equation}
This can be solved for $\sigma_c$:
\begin{equation}
\begin{array}{ll}
R_{XY}^2  = 1 / \left [ \left ( 1 + (\sigma_{sX}/\sigma_c)^2 \right ) \left ( 1 + (\sigma_{sY}/\sigma_c)^2 \right ) \right ] \\
\Rightarrow R_{XY}^2 \left [ 1 + (\sigma_{sX}/\sigma_c)^2 + (\sigma_{sY}/\sigma_c)^2 + (\sigma_{sX}/\sigma_c)^2 (\sigma_{sY}/\sigma_c)^2 \right ] = 1 \\
\Rightarrow \left [ \sigma_c^4 + \sigma_c^2(\sigma_{sX}^2 + \sigma_{sY}^2) + \sigma_{sX}^2 \sigma_{sY}^2 \right ] = \sigma_c^4 / R_{XY}^2 \\
\Rightarrow \sigma_c^4(1 - 1/R_{XY}^2)  + \sigma_c^2(\sigma_{sX}^2 + \sigma_{sY}^2) + \sigma_{sX}^2 \sigma_{sY}^2 = 0 \\
\end{array}
\end{equation}
This is a quadratic equation in $\sigma_c^2$ whose solution is:
\begin{equation}
\begin{array}{ll}
\sigma_c^2 = A \pm B; \\
A = - (\sigma_{sX}^2 + \sigma_{sY}^2) / C; \\
B = \left [ (\sigma_{sX}^2 + \sigma_{sY}^2)^2 - 4 (1 - 1/R_{XY}^2) \sigma_{sX}^2 \sigma_{sY}^2 \right ]^{1/2} / C \\
C = 2 (1 - 1/R_{XY}^2). \\
\end{array}
\end{equation}
One of the solutions is negative which is unphysical, so we adopt the positive solution. 

Now the mean counts in the P1, P2 and P3 phases in Fig.~\ref{fig3} are $80.7$, $100.8$ 
and $72.2$ respectively. The corresponding $\sigma_{s1}$, $\sigma_{s2}$ and $\sigma_{s3}$ 
are $0.1113$, $0.0996$ and $0.1177$ respectively. These are of the order of $0.1$ while 
the rms of the correlated signal $\sigma_c$ is $\approx 0.007$, which is much smaller. 
Note that all these values are relative to the mean counts in that phase. In this limit 
the cross correlation simplifies to
\begin{equation}
R_{XY} \approx \sigma_c^2 / \left (\sigma_{sX} \sigma_{sY} \right )
\end{equation}

\section{``Ecological fallacy"}

\begin{figure}
\begin{center}
\advance\leftskip-0.3cm
\includegraphics[keepaspectratio=true,scale=1.0,width=9.5cm]{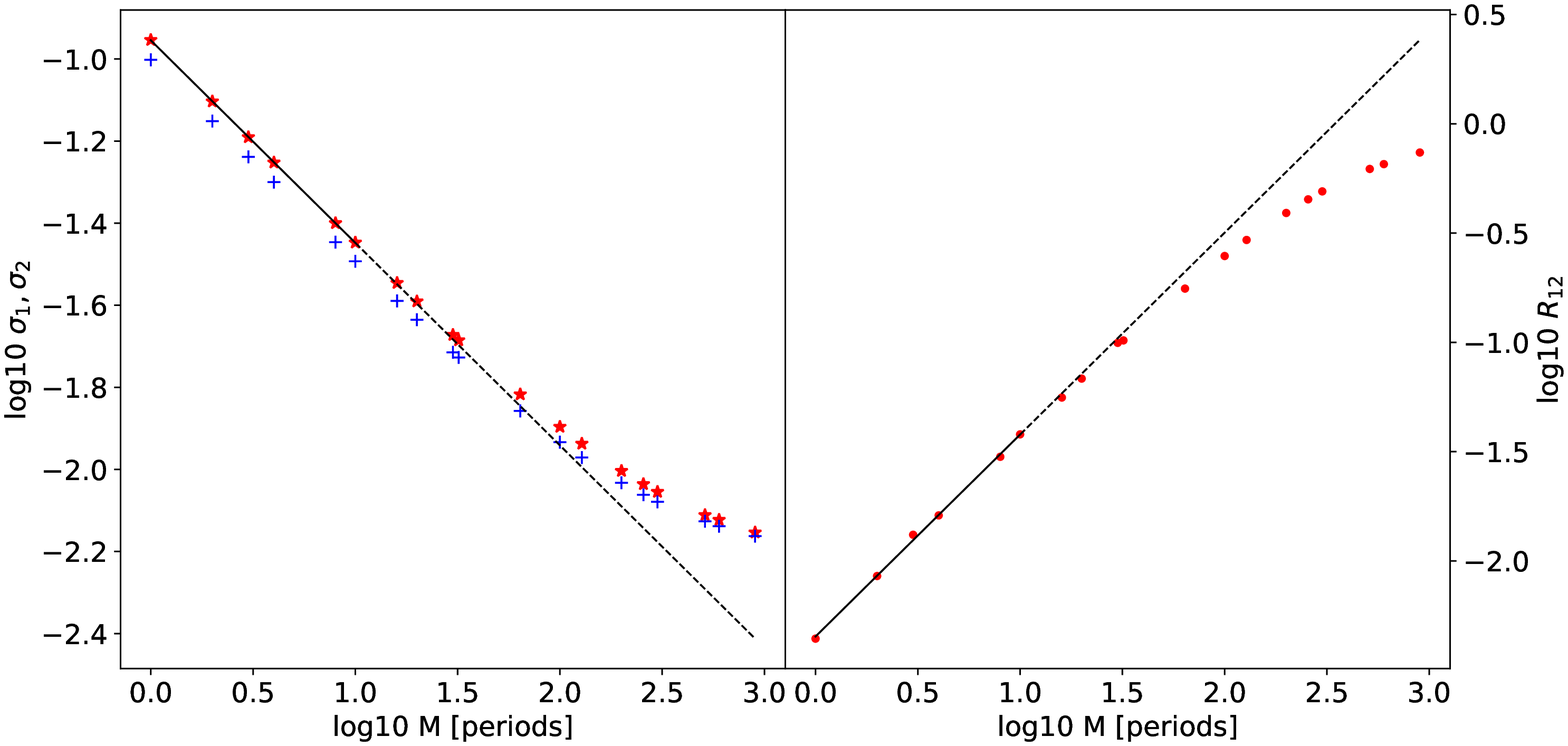}
\end{center}
\vskip-0.5cm
\caption{
	(a) Left panel: Estimated standard deviations of the data in phases P1 
	(red stars) and P2 (blue pluses) respectively, as a function of the number 
	of periods averaged ($M$). The straight line is the least squares fit to 
	the first $6$ data of P1; the dotted line is its continuation. (b) Right 
	panel: The cross correlation $R_{12}$ of the averaged data as a function 
	of $M$.. The straight line is the least squares fit to the first $6$ data; 
	the dotted line is its continuation.
        }
\label{fig5}
\end{figure}

Consider what happens when the data of Appendix A is averaged before cross correlation,
say by $M$ samples. Now one has a new data set $U_i, V_i$ ($i = 1, ..., N/M$, assuming
that $N$ is an integral multiple of $M$), where
\begin{equation}
\begin{array}{ll}
U_i  = (\sum_{j=k}^l X_j) / M; V_i  = (\sum_{j=k}^l Y_j) / M;  \\
k = (i - 1) \times M + 1; l = i \times M. \\
\end{array}
\end{equation}
The correlation coefficient $R_{UV}$ will be approximately $R_{XY}\times M$ as long as
the time scale of variation of the correlated signal $c$ is much larger than $M$ samples;
then the numerator in equation A.1 will not change while the denominator decreases as 
$1/M$. This is evident in equation~A.9.

The left panel of Fig.~\ref{fig5} shows the variation of the estimated standard 
deviations $\sigma_1$ and  $\sigma_2$ of the data in phases P1 and P2 respectively, as a 
function of the number of periods averaged ($M$), in a log10-log10 plot. These are the 
$\sigma_X$ and $\sigma_Y$ used in equation~A.1, and should be equivalent to 
$\sqrt{\sigma_c^2 + \sigma_{sX}^2}$ and $\sqrt{\sigma_c^2 + \sigma_{sY}^2}$ respectively. 
Initially both  parameters decrease inversely with $\sqrt{M}$ as expected, then the 
decrease is slower. A least squares straight line fit to the first six data of each 
standard deviation yields the results shown in Table~\ref{tbl4}.

\begin{table}
\begin{center}
\caption{
	Slope and intercept of the straight lines fit in a least squares sense to
	the logarithm (of base $10$) of $\sigma_{1}$ and $\sigma_{2}$ as a function 
	of logarithm of $M$.
        } \label{tbl4}
\begin{tabular}{|c|c|c|}
\hline
\hline
 Phase & Slope & Intercept\\
\hline
\hline
 P1 & $ -0.4935(9)$ & $-0.9545(6)$ \\
\hline
 P2 & $-0.4908(14)$ & $-1.0033(9)$ \\
\hline
\hline
\end{tabular}
\end{center}
\end{table}

Initially the slopes are very close to the expected value $-0.5$, then the slope
reduces in magnitude. Now, the observed value of $\sigma_{1}$ at $M = 900$ (the
last red point in the left panel of Fig.~\ref{fig5}) is $0.0070$  while the
value expected on the basis of the straight line is $0.0039$. Since the standard
deviations add quadratically, the difference rms is $\sqrt{0.0070^2 - 0.0039^2} 
= 0.0058$ which is quite close to the expected value of $\approx 0.0070$. In 
other words the red stars data is the square root of the quadratic addition of 
a fixed rms of about $0.0058$ and a variable rms of $\approx 0.1/\sqrt{M}$.
Similar results occur for $\sigma_{2}$ in the figure. It is therefore concluded 
that $\sigma_{1}$ and $\sigma_{2}$ vary as expected in Fig.~\ref{fig5}. In the 
right panel of Fig.~\ref{fig5} $R_{12}$ varies correspondingly. A least squares 
straight line fit to the first six data give slope $= 0.9239(5)$ and intercept 
$= -2.3458(3)$. Initially the correlation varies almost linearly with $M$ as 
expected, then varies more slowly. The intercept implies a correlation of 
$10^{-2.3458} = 0.0045$ which is consistent with the $R_{12}$ of Table~\ref{tbl1}.

Fig.~\ref{fig5} is a check on the basic health of our data.

\section{Details of filtering data}

Let $g1$, $g2$ and $g3$ be the X-ray fluxes in the three phase components P1, P2
and P3 respectively; each contains $\approx 4.06$ million samples. Let their mean
values be $<g1>$, $<g2>$ and $<g3>$. Let $h1$, $h2$ and $h3$ be the smooth versions
of $g1$, $g2$ and $g3$ respectively, obtained using a smoothing window of $N_{sw}$ 
samples and one of the three filtering algorithms. Then two new data are formed as 
follows: 
\begin{equation}
\begin{array}{ll}
g1^\prime = g1 - ( <g1> / <g3> ) h3; & g1^{\prime\prime} = g1 - h1 \\
g2^\prime = g2 - ( <g2> / <g3> ) h3; & g2^{\prime\prime} = g2 - h2 \\
g3^\prime = g3 - h3;                 & g3^{\prime\prime} = g3^\prime \\
\end{array}
\end{equation}

The primed data is the original data from which the normalized version of $h3$ is 
subtracted, the normalization constant being equal to the ratio of the means of the
data. The double primed data is the original data from each of which the corresponding 
smooth data is subtracted. The correlation of the primed data is expected to contain
information about correlated current variations in the on pulse, and is represented
by $R^\prime$; correlation of the double primed data is expected to be zero since it
should essentially be white noise, and is represented by $R^{\prime\prime}$. However
due to practical reasons $R^{\prime\prime}$ will have a small but finite value; this
will act as a calibrator for $R^\prime$. The information we seek lies in the quantity 
$R^\prime - R^{\prime\prime}$, since correlations of small values add linearly; it is
this quantity that is listed as $R$ in Table~\ref{tbl2}.

\begin{figure}
\begin{center}
\advance\leftskip-0.3cm
\includegraphics[keepaspectratio=true,scale=1.0,width=8.5cm]{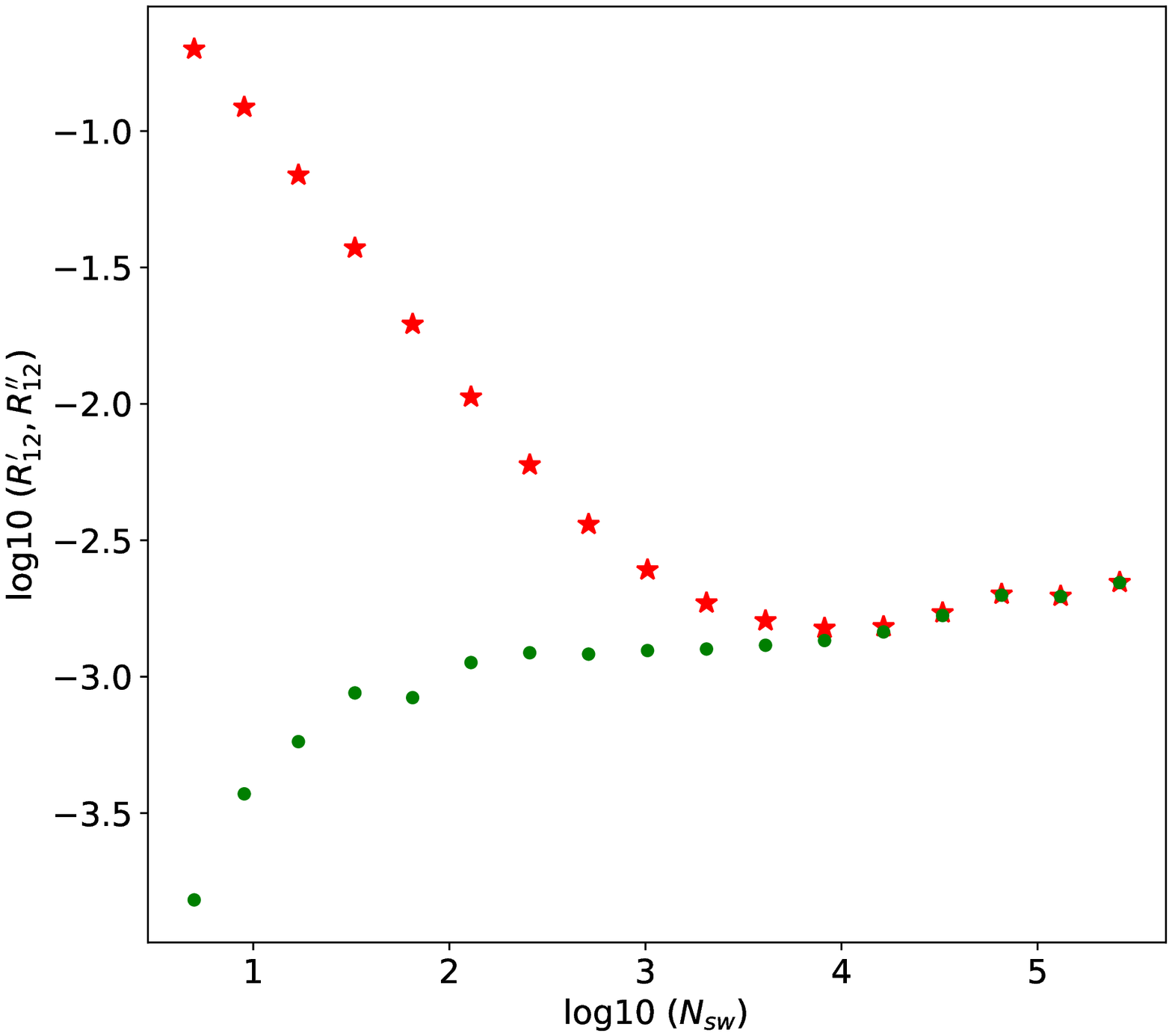}
\end{center}
\vskip-0.5cm
\caption{
	$R_{12}^\prime$ (red stars) and $R_{12}^{\prime\prime}$ (green dots) plotted 
	against the smoothing window width $N_{sw}$, both in the log10 scale, for 
	the SG filter.
        }
\label{fig6}
\end{figure}

Figure~\ref{fig6} shows the variation of $R_{12}^\prime$ and $R_{12}^{\prime\prime}$ 
with $N_{sw}$ for the SG filter; these correspond to cross correlation of data of the
phase components P1 and P2. $R_{12}^{\prime\prime}$ values show some variation
but are all relatively small values as expected; these represent zero correlation for 
each $N_{sw}$. On the other hand $R_{12}^\prime$ shows a relatively significant 
variation, tending to its true value in an almost asymptotic manner as $N_{sw}$ 
increases. This is also expected since a larger $N_{sw}$ would imply better estimation 
of the smooth functions $h1$ and $h2$.

\begin{figure}
\begin{center}
\advance\leftskip-0.3cm
\includegraphics[keepaspectratio=true,scale=1.0,width=8.5cm]{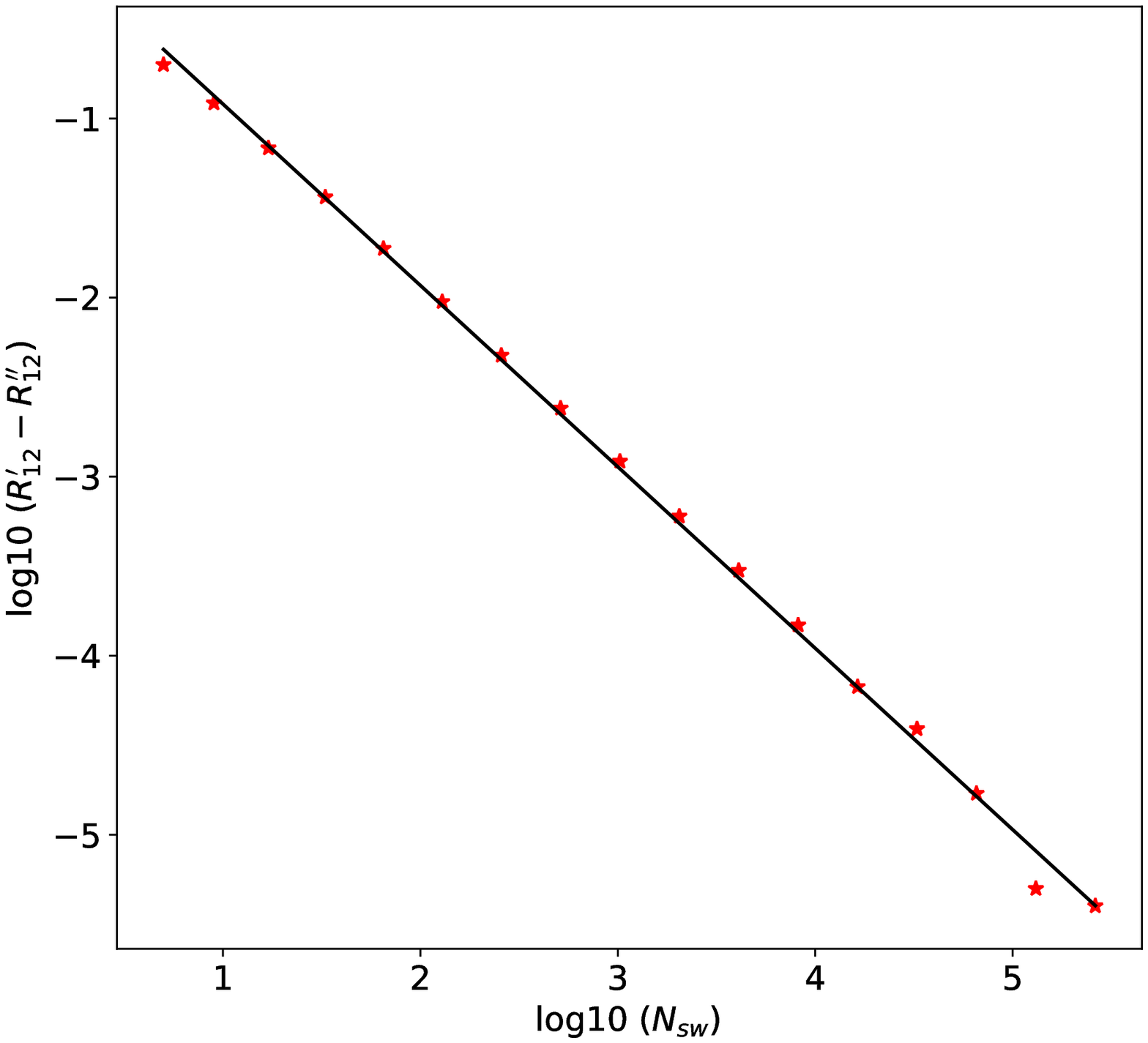}
\end{center}
\vskip-0.5cm
\caption{
	The difference $R_{12}^\prime - R_{12}^{\prime\prime}$ plotted against the 
	$N_{sw}$, both in the log10 scale, for the SG filter.
        }
\label{fig7}
\end{figure}

Figure~\ref{fig7} shows the difference correlation $R_{12}^\prime - R_{12}^{\prime\prime}$ 
plotted against $N_{sw}$ for the SG filter. A least squares straight line fit yields the
slope $-0.97 \pm 0.01$ and intercept $0.03 \pm 0.02$. Clearly this difference appears to
decrease inversely with $N_{sw}$ so it tends to the value $0.0$ at large $N_{sw}$. So we 
can only use the smallest value of the difference (last data in Figure~\ref{fig7}) as an
upper limit for the correlation in Table~\ref{tbl2}. The errors in Table~\ref{tbl2} are 
derived from the last three data in Figure~\ref{fig7}, three being the minimum number of
data required to estimate the standard deviation. Plots of $R_{12}^\prime - 
R_{12}^{\prime\prime}$ for the other two filters also show a similar trend with $N_{sw}$.

Similar plots have been made for the differences $R_{13}^\prime - R_{13}^{\prime\prime}$
and $R_{23}^\prime - R_{23}^{\prime\prime}$ for all three filters. These show a damped 
oscillation kind of trend around the value $0.0$, so they are also tending to the value 
$0.0$ at large $N_{sw}$. The smallest difference of these plots has been used as upper 
limit for the correlations in Table~\ref{tbl2}, as was done above, including for their
errors.

In summary this section shows that there is almost no difference between $R^\prime$ and 
$R^{\prime\prime}$ as the smoothing filter estimates the smoothing function better by
an increase of $N_{sw}$.


\begin{thebibliography}{}
\bibitem[Arons (1983)]{Arons1983} Arons, J. 1983, \apj, 266, 215
\bibitem[Arons (2009)]{Arons2009} Arons, J. 2009, Astrophysics and Space Science Proceedings "Neutron Stars and Pulsars", Eds. Werner Becker, pp 373, Springer.
\bibitem[Arons (2011)]{Arons2011} Arons, J. 2011, Astrophysics and Space Science Proceedings "High-Energy Emission from Pulsars and their Systems", Proceedings of the First Session of the Sant Cugat Forum on Astrophysics, Eds. Nanda Rea \& Diego F. Torres, pp 165, Springer.
\bibitem[Arons \& Scharlemann (1979)]{Arons1979} Arons, J. \& Scharlemann, E.T. 1979, \apj, 231, 854
\bibitem[Bowley (1928)]{Bowley1928} Bowley, A.L 1928, Journal of the American Statistical Association, Vol. 23, No. 161, pp 31; https://doi.org/10.2307/2277400
\bibitem[Cerutti et al. (2016)]{Cerutti2016} Cerutti, B., . Philippov, A.A. and Spitkovsky, A. 2016, \mnras, 457, 2401
\bibitem[Cheng (2011)]{Cheng2011} Cheng, K.S. 2011, Astrophysics and Space Science Proceedings "High-Energy Emission from Pulsars and their Systems", Proceedings of the First Session of the Sant Cugat Forum on Astrophysics, Eds. Nanda Rea \& Diego F. Torres, pp 99, Springer.
\bibitem[Cheng et al. (1986)]{Cheng1986} Cheng, K.S., Ho, C. \& Ruderman, M. 1986, \apj, 300, 500
\bibitem[Contopoulos et al. (1999)]{Contopoulos1999} Contopoulos, I., Kazanas, D., Fendt, C. 1996, \apj, 511, 351
\bibitem[Contopoulos et al. (2019)]{Contopoulos2019} Contopoulos, I., Petri, J., Stefanou, P. 2019, \mnras, 491, 5589
\bibitem[Goldreich \& Julian  (1969)]{Goldreich1969} Goldreich, P. \& Julain, W.H. 1969, \apj, 157, 869
\bibitem[Harding \& Grenier (2011)]{Harding2011} Harding, A.K. \& Grenier, A. 2011, Astrophysics and Space Science Proceedings "High-Energy Emission from Pulsars and their Systems", Proceedings of the First Session of the Sant Cugat Forum on Astrophysics, Eds. Nanda Rea \& Diego F. Torres, pp 79, Springer.
\bibitem[Harding (2022)]{Harding2022} Harding, A.K. 2022, Astrophysics and Space Science Library 465 "Millisecond Pulsars", Eds. Sudip Bhattacharyya, Alessandro Papitto \& Dipankar Bhattacharya, pp 57, Springer.
\bibitem[Hermsen et al. (2013)]{Hermsen2013} Hermsen, W., Hessels, J.W.T., Kuiper, L. et al. 2013, Science 339, 436
\bibitem[Hermsen et al. (2017)]{Hermsen2017} Hermsen, W., Kuiper, L., Hessels, J.W.T. et al. 2017, \mnras 466, 1688
\bibitem[Hermsen et al. (2018)]{Hermsen2018} Hermsen, W., Kuiper, L., Basu, R. et al. 2018, \mnras, 480, 3655
\bibitem[Hirotani (2011)]{Hirotani2011} Hirotani, K. 2011, Astrophysics and Space Science Proceedings "High-Energy Emission from Pulsars and their Systems", Proceedings of the First Session of the Sant Cugat Forum on Astrophysics, Eds. Nanda Rea \& Diego F. Torres, pp 117, Springer.
\bibitem[Kirk et al (2009)]{Kirk2009} Kirk, J.G., Lyubarsky, Y. \& Petri, J. 2009, Astrophysics and Space Science Proceedings "Neutron Stars and Pulsars", Eds. Werner Becker, pp 421, Springer.
\bibitem[Mereghetti et al. (2016)]{Mereghetti2016} Mereghetti1, S., Kuiper, L., Tiengo, A. et al. 2016, \apj, 831, 21
\bibitem[Ostriker, \& Gunn  (1969)]{Ostriker1969} Ostriker, J.P. \& Gunn, J.E. 1969, \apj, 157, 1395
\bibitem[Petri (2011)]{Petri2011} Petri, J. 2011, Astrophysics and Space Science Proceedings "High-Energy Emission from Pulsars and their Systems", Proceedings of the First Session of the Sant Cugat Forum on Astrophysics, Eds. Nanda Rea \& Diego F. Torres, pp 181, Springer.
\bibitem[Philippov et al. (2020)]{Philippov2020} Philippov, A., Timokhin, A., Spitkovsky, A. 2020, \prl, 124, 245101
\bibitem[Ruderman \& Sutherland (1975)]{Ruderman1975} Ruderman, M.A. \& Sutherland, P.G. 1975, \apj, 196, 51
\bibitem[Sobey et al. (2015)]{Sobey2015} Sobey, C., Young, N.J., Hessels, J.W.T. 2015, \mnras, 451, 2492
\bibitem[Spitkovsky (2006)]{Spitkovsky2006} Spitkovsky, A. 2006, \apj, 648, L51
\bibitem[Spitkovsky (2011)]{Spitkovsky2011} Spitkovsky, A. 2011, Astrophysics and Space Science Proceedings "High-Energy Emission from Pulsars and their Systems", Proceedings of the First Session of the Sant Cugat Forum on Astrophysics, Eds. Nanda Rea \& Diego F. Torres, pp 139, Springer.
\bibitem[Sturrock (1971)]{Sturrock1971} Sturrock, P.A. 1971, \apj, 164, 529
\bibitem[Tennant et al. (2001)]{Tennant2001} Tennant, A.F., Becker, W., Juda, M. et al. 2001, \apj, 554, L173
\bibitem[Tuo et al. (2019)]{Tuo2019} Tuo, Y.L., Ge, M.Y., Song, L.M. et al. 2019, RAA, 19, 87
\bibitem[Vivekanand (2020)]{Vivekanand2020} Vivekanand, M. 2020, \aap, 633, A57
\bibitem[Vivekanand (2021)]{Vivekanand2021} Vivekanand, M. 2021, \aap, 649, A140
\bibitem[Vivekanand (2022)]{Vivekanand2022} Vivekanand, M. 2022, \mnras, 514, 185
\end{thebibliography}
\end{document}